\documentclass[aps, twocolumn, showpacs, letterpaper]{revtex4}

\usepackage{amsmath}
\usepackage{amssymb}
\usepackage{graphicx}
\usepackage{xspace}
\usepackage{sbraket}

\newcommand{\eg}{{e.g.,\/}\xspace}
\newcommand{\ie}{{i.e.,\/}\xspace}

\newcommand{\eq}[1]{(\ref{#1})}
\newcommand{\Eq}[1]{Eq.~(\ref{#1})}
\newcommand{\Eqs}[1]{Eqs.~(\ref{#1})} 
\newcommand{\Eqsc}[2]{Eqs.~(\ref{#1}), (\ref{#2})}
\newcommand{\Eqsd}[2]{Eqs.~(\ref{#1})-(\ref{#2})}
\newcommand{\Sec}[1]{Sec.~\ref{#1}}
\newcommand{\App}[1]{Appendix~\ref{#1}}
\newcommand{\Fig}[1]{Fig.~\ref{#1}} 
\newcommand{\Ref}[1]{Ref.~\cite{#1}}

\newcommand{\mc}[1]{\mathcal{#1}}
\newcommand{\mcc}[1]{\mathfrak{#1}}

\newcommand{\kpt}[1]{{\kern #1 pt}}

\renewcommand{\vec}[1]{{\boldsymbol{\rm #1}}}
\newcommand{\oper}[1]{\hat{#1}}
\newcommand{\const}{\text{const}}
\renewcommand{\Re}{\text{Re}\,}

\newcommand{\rind}[1]{{\scriptscriptstyle #1}}
\newcommand{\transp}{^{\text{T}}}
\newcommand{\sumprime}{\kpt{-4}^{'}\kpt{2}}
\newcommand{\twobytwomatrix}[4]{\left(\begin{array}{cc}#1 & #2\\#3 & #4\end{array}\right)}

\newcommand{\action}{\mc{J}}
\newcommand{\freq}{\varpi}
\newcommand{\vaction}{\vec{\action}}
\newcommand{\vfreq}{\vec{\freq}}
\newcommand{\vxi}{\vec{\xi}}
\newcommand{\vpi}{\vec{\pi}}
\newcommand{\vxie}{\bar{\vxi}}
\newcommand{\dright}{r}
\newcommand{\dleft}{\ell}
\newcommand{\dbright}{\ket{\dright}}
\newcommand{\dbleft}{\bra{\dleft}}
\newcommand{\meff}{m_\text{eff}}
\newcommand{\rabi}{\Omega_{\rm R}}
\newcommand{\lint}{\tilde{\mc{L}}_{\rm int}}
\newcommand{\potl}{\mc{U}}
\newcommand{\lmode}{\tilde{\mc{L}}}
\newcommand{\dfreq}{\delta\kpt{-.7}\nu\kpt{.2}}
\newcommand{\deltam}{\delta\kpt{-.3}m\kpt{.8}}
\newcommand{\dxi}{\kpt{.5}\dot{\kpt{.5}\vxi\kpt{.5}}}

\begin{document}

\title{Dressed-particle approach in the nonrelativistic classical limit}
\author{I. Y. Dodin and N. J. Fisch}
\affiliation{Department of Astrophysical Sciences, Princeton University, Princeton, New Jersey 08544, USA}
\date{\today}

\begin{abstract}
For a nonrelativistic classical particle undergoing arbitrary oscillations, the generalized effective potential $\Psi$ is derived from nonlinear eigenfrequencies of the particle-field system. Specifically, the ponderomotive potential is extended to a nonlinear oscillator, resulting in multiple branches near the primary resonance. For a pair of natural frequencies in a beat resonance, $\Psi$ scales linearly with the internal actions and is analogous to the dipole potential for a two-level quantum system. Thus cold quantum particles and highly-excited quasiclassical objects permit uniform manipulation tools, particularly, one-way walls.
\end{abstract}



\maketitle

\section{Introduction} 

Multiscale adiabatic dynamics of classical particles in oscillating and static fields is simplified within the oscillation-center (OC) approach, which allows separating fast quiver motion of the particles from their slow translational motion \cite{ref:dewar73, ref:similon86, my:dipole}. Hence the average forces are embedded into the properties of the OC, yielding a quasiparticle with a variable effective mass $\meff$ \cite{my:mneg, proc:dodin08}. In each given case, $\meff$ can be Taylor-expanded at nonrelativistic energies so as to appear as an effective potential $\Psi$ \cite{my:mneg}, \eg ponderomotive \cite{ref:gaponov58, ref:motz67, ref:cary77, ref:hatori81} or diamagnetic \cite{book:jackson}. Yet the nonrelativistic limit must permit also an independent calculation of $\Psi$. For linear oscillations, the generalized effective potential was derived in \Ref{my:dipole}. However, a comprehensive method of finding $\Psi$ for nonlinear quiver motion has not been proposed.

The purpose of this work is to calculate, from first principles, the generalized effective potential $\Psi$ for a nonrelativistic classical particle undergoing arbitrary oscillations in high-frequency or static fields. We proceed by finding eigenmodes in the particle-field system; hence $\Psi$ is obtained like in the dressed-atom approach \cite{book:cohentannoudji, ref:dalibard85, ref:berman82, ref:courtens77} but from nonlinear classical equations. Specifically, we show that the ponderomotive potential extended to a nonlinear oscillator has multiple branches near the primary resonance. Also, for a pair of natural frequencies in a beat resonance, $\Psi$ scales linearly with the internal actions and is analogous to the dipole potential for a two-level quantum system.  Thus cold quantum particles and highly-excited quasiclassical objects permit uniform manipulation tools, particularly, stationary asymmetric barriers, or one-way walls \cite{my:cdlarge, my:ratchet, ref:raizen05, ref:ruschhaupt04, ref:ruschhaupt06, ref:ruschhaupt06c, ref:price07, ref:price08, ref:thorn08}. 

The work is organized as follows. In \Sec{eq:basic}, we obtain the general form of the effective potential $\Psi$. In \Sec{sec:eigen}, we derive the equations for oscillation modes. In \Sec{sec:pond}, we calculate the ponderomotive potential from the infinitesimal frequency shift of the oscillating field coupled to a particle at a primary resonance, both linear and nonlinear; see also \App{app:ponder}. In \Sec{sec:hyb}, we find $\Psi$ near a beat resonance and show the analogy with the dipole potential. In \Sec{sec:oneway}, we explain how $\Psi$ allows one-way walls. In \Sec{sec:conclusions}, we summarize our main results.

\section{Generalized effective potential} 
\label{eq:basic}

Consider a particle which exhibits slow dynamics in canonical variables $(\vec{r},\vec{P})$ superimposed on fast oscillations in angle-action variables $(\vec{\vartheta},\vaction=\const)$, such that zero $\vaction$ corresponds to purely translational motion. The OC Lagrangian is then written as \cite{my:mneg}
\begin{gather}
\mc{L}_0=-\meff c^2\sqrt{1-v^2/c^2},
\end{gather}
where $\vec{v} \equiv \dot{\vec{r}}$, and $\meff(\vec{r},\vec{v}; \vaction)$ is the effective mass, the dependence on $\vaction$ being parametric \cite{foot:time}. Thus the complete Lagrangian, which describes also the oscillations, equals $\mc{L}_0 + \dot{\vec{\vartheta}}\cdot\vaction$, so the corresponding Hamiltonian $\mc{H}$ matches that of the oscillation center: $\mc{H}=\vec{P}\cdot\vec{v}-\mc{L}_0$.

Assume nonrelativistic dynamics, \ie $v\ll c$ and $\deltam \equiv {\nobreak\meff - m} \ll m$, where $m$ is the true mass. Hence
\begin{gather}
\mc{L}_0=\frac{1}{2}\,mv^2-\potl, \quad \potl = \deltam c^2,
\end{gather}
so the Hamiltonian reads
\begin{gather}
\mc{H}=\frac{P^2}{2m} + \Psi, \quad \Psi = \potl -\frac{1}{2m}(\partial_\vec{v}\potl)^2,
\end{gather}
where $\vec{P} = m\vec{v} - \partial_\vec{v}\potl$, and $\Psi(\vaction=0)=0$. 

Following \Ref{my:mneg}, one can find $\Psi$ from the relativistic particle trajectory in given fields; however, a general nonrelativistic approach is also possible. Formally, field modes can be understood as particle degrees of freedom. Then, like in the dressed-atom approach \cite{book:cohentannoudji, ref:dalibard85}, the effective potential can be found from eigenfrequencies $\vfreq \equiv \dot{\vec{\vartheta}} = \partial_\vaction\mc{H}$ of the particle-field system:
\begin{gather}\label{eq:psi}
\Psi = \int \vfreq\cdot d\vaction.
\end{gather}
In this case, $\Psi$ depends on $\vec{r}$ and $\vec{P}$ only parametrically, through $\vfreq(\vec{r},\vec{P},\vaction)$.

The canonical frequencies can be redefined such that $\vfreq \to \vfreq +\kpt{1} \const$, adding a constant to $\Psi$. Albeit arbitrarily large, this contribution does not affect the motion equations, so we abandon the requirement that $\Psi$ must remain small compared to $mc^2$; hence actual physical frequencies can be used for $\vfreq$. Particularly, for uncoupled modes one gets $\Psi=\Psi_0$, 
\begin{gather}\label{eq:psi0}
\Psi_0 = \vec{\Omega}\cdot\vec{J} + \vec{\omega}\cdot\vec{I},
\end{gather}
where we used \Eq{eq:psi} and $(\vec{\Omega},\vec{J})$, $(\vec{\omega},\vec{I})$ for the unperturbed frequencies and actions of the particle and the field, correspondingly. For an unbounded field ($I \to \infty$), the second term in \Eq{eq:psi} is infinite. As it is fixed though, the force on a particle is determined only by $\vec{\Omega}\cdot\vec{J}$ and the finite modification of the effective potential due to coupling,
\begin{gather}\label{eq:phipsi0}
\Phi \equiv \Psi - \Psi_0,
\end{gather}
where $\Phi$ also can be found from \Eq{eq:psi}, as shown below.

\section{Partial mode decomposition} 
\label{sec:eigen}

Suppose weakly nonlinear oscillations $\vxi(t)$, both of the particle \cite{my:dipole} and of external fields \cite{foot:fieldmode}, so their Lagrangian reads $\lmode(\vxi,\dxi) = \lmode_0 + \lint$, where $\lint$ is a perturbation to a bilinear form $\lmode_0$ \cite{foot:dipole},
\begin{gather}\label{eq:l0}
\lmode_0 = 
\frac{1}{2}\,(\dxi\cdot\oper{M}\dxi)-
(\dxi\cdot\oper{R}\vxi)-
\frac{1}{2}\,(\vxi\cdot\oper{Q}\vxi).
\end{gather}
Here $\oper{M}$, $\oper{R}$, $\oper{Q}$ are $N\times N$ real matrices; $\oper{M}$ and $\oper{Q}$ are symmetric, $\oper{R}$ is antisymmetric, and $\text{rank}\,\oper{M} = N \equiv \text{dim}\,\vxi$. At zero $\oper{R}$, $\lmode_0(\vxi,\dxi)$ can be diagonalized to yield
\begin{gather}\label{eq:decompose1}
\lmode_0 = \sum_{j=1}^N L_j, \quad L_j = \frac{1}{2}\,M_j\dot{\xi}_j^2-\frac{1}{2}\,Q_j\xi_j^2,
\end{gather}
where $L_j$ describe individual modes $\xi_j$ \cite{book:goldstein}. Then
\begin{gather}\label{eq:reald}
\oper{D}_j\xi_j= \delta_{\xi_j}\lint, \quad 
\oper{D}_j = M_j\,d^{\kpt{.9}2}_t+Q_j,
\end{gather}
$\delta$ and $d_t$ standing for the variational and time derivatives. Yet, such decomposition does not hold in the general case, so we redefine eigenmodes, following \Ref{ref:schenk01}.

Extend the configuration space by introducing
\begin{gather}
\dbleft =(-\vpi\oper{M}^{-1}, \vxi),\quad
\dbright = (\vxi,\oper{M}^{-1}\vpi)\transp
\end{gather}
as the new, ``left'' and ``right'', coordinate vectors, where $\vpi = \oper{M}\dxi-\oper{R}\vxi$ is the old canonical momentum. Then
\begin{gather}\label{eq:lagrs}
\lmode_0 =
\frac{1}{4}\,\big[\braket{\dot{\dleft}|\oper{\mcc{M}}|\dright}-\braket{\dleft|\oper{\mcc{M}}|\dot{\dright}}\big]
+\frac{1}{2}\,\braket{\dleft|\oper{\mcc{F}}|\dright},
\end{gather}
where we omitted a full time derivative and introduced
\begin{gather}
\oper{\mcc{M}}=\twobytwomatrix{\oper{M}}{0}{0}{\oper{M}}, \quad
\oper{\mcc{F}} = \twobytwomatrix{\oper{R}}{\oper{M}}{\oper{F}}{\oper{R}}, 
\end{gather}
with $\oper{F} = \oper{R}\oper{M}^{-1}\!\oper{R}-\oper{Q}$. Thus the resulting equations are
\begin{gather}\label{eq:eigens}
\bra{\dot{\dleft}}\oper{\mcc{M}} + \dbleft \oper{\mcc{F}}  = 0,\quad
\oper{\mcc{M}}\ket{\dot{\dright}} - \oper{\mcc{F}} \dbright = 0, 
\end{gather}
both equivalent to
\begin{gather}\label{eq:eqxi}
\oper{M}\ddot{\vxi}-2\oper{R}\dxi + \oper{Q}\vxi = 0.
\end{gather}

\Eq{eq:eqxi} has $2N$ eigenmodes $\vxi_j = \vxie_j\,e^{-i\nu_j t}$, with $\nu_j$ hence assumed real and nonzero; therefore, for each $\vxi_j$, there also exists a mode $\vxi_{-j} = \vxi_j^*$, and $\vxie_j$ are generally not orthogonal. The corresponding eigenmodes of \Eqs{eq:eigens} are
\begin{gather}
\bra{\dleft_j} = e^{i\nu_j t}\bra{\bar{\dleft}_j}, \quad
\ket{\dright_j} = e^{-i\nu_j t} \ket{\bar{\dright}_j},
\end{gather}
with vector amplitudes
\begin{gather}
\bra{\bar{\dleft}_j} =(-i\nu_j\vxie^*_j-\vxie_j^*\oper{R}\oper{M}^{-1},\vxie_j^*),\\
\ket{\bar{\dright}_j} = (\vxie_j, -i\nu_j\vxie_j-\oper{M}^{-1}\oper{R}\vxie_j)\transp,
\end{gather}
and $\mcc{M}_{jk}\equiv \braket{\bar{\dleft}_j|\oper{\mcc{M}}|\bar{\dright}_k}=-2i\rho_{jk}$, where
\begin{gather}\label{eq:ort}
\rho_{jk} = \frac{1}{2}\, \,\vxie_j^*\cdot\big[(\nu_j+\nu_k)\oper{M}-2i\oper{R}\big]\cdot\vxie_k.
\end{gather}
The matrix $\oper{\rho}$ is diagonal for distinct $\nu_j$, as seen from \Eq{eq:eqxi}, or can be diagonalized when some of the frequencies coincide \cite{ref:schenk01}; thus, 
\begin{gather}
\mcc{M}_{jk} = -2i \rho_j\delta_{jk},\quad
\rho_j \equiv \nu_j (\vxie_j^*\cdot\oper{M}\vxie_j),
\end{gather}
$\rho_j=-\rho_{-j}$. (Hence modes with $\nu_j=0$ are orthogonal to the others and can be considered separately, as implied below.) Therefore any $\dbleft$ and $\dbright$ are decomposed as
\begin{gather}
\dbleft = \sum_{j=-N}^{N}\sumprime \dleft_j\bra{\bar{\dleft}_j},\quad
\dbright = \sum_{j=-N}^{N}\sumprime \dright_j\ket{\bar{\dright}_j},\label{eq:expansion}
\end{gather}
where the primes stand for skipping $j=0$, and
\begin{gather}
\dleft_j = \frac{i}{2\rho_j}\,\braket{\dleft|\oper{\mcc{M}}|\bar{\dright}_j}, \quad
\dright_j = \frac{i}{2\rho_j}\,\braket{\bar{\dleft}_j|\oper{\mcc{M}}|\dright}.\label{eq:expand1}
\end{gather}

Since $\dbleft$ and $\dbright$ are real, one has $\dright_j = \dleft_j^*\equiv \psi_j\sqrt{2}$ and $\psi_{-j} = \psi_j^*$; hence \Eq{eq:decompose1}, but with
\begin{gather}\label{eq:compagr}
L_j = 
\frac{i\rho_j}{2}\left(\dot{\psi}_j\psi_j^*-\psi_j\dot{\psi}_j^*\right)-\rho_j\nu_j |\psi_j|^2.
\end{gather}
The resulting equations for individual modes are
\begin{gather}\label{eq:response}
\oper{D}_j \psi_j = \delta_{\psi_j^*} \lint, \quad 
\oper{D}_j = \rho_j\left(\nu_j-i\kpt{.5}d_t\right),
\end{gather}
similar to reduced \Eqs{eq:reald}. Particularly, at zero $\lint$,
\begin{gather}
\psi_j=\sqrt{2\action_j}\,e^{-i\vartheta_j}, \quad 
\dot{\vartheta}_j = \nu_j, \quad 
\action_j = \const.
\end{gather}
On the other hand, $L_j = (\dot{\vartheta}_j-\nu_j)\action_j$; thus $\partial_{\dot{\vartheta}_j}L_j = \action_j$ is also the action corresponding to the angle $\vartheta_j$:
\begin{gather}\label{eq:action}
\action_j = \rho_j |\psi_j|^2,
\end{gather}
so $\nu_j$ is the canonical frequency. Then the mode energy is $\nu_j\action_j$ (thus $\rho_{j}>0$ for stable modes with $\nu_j>0$, henceforth implied), and \Eq{eq:psi0} is recovered.

In the next sections, we apply \Eqs{eq:response} to find eigenmodes for nonzero $\lint$, with $\vec{\psi}$ becoming partial oscillations. Hence the effective potential modification $\Phi$ [\Eq{eq:phipsi0}] is obtained from \Eq{eq:psi}.

\section{Primary resonance} 
\label{sec:pond}

\subsection{Linear oscillator}
\label{sec:pondlin}

First, we calculate $\Psi$ for a linear coupling between a pair of modes $\psi_1$ and $\psi_2$, say, 
\begin{gather}
\lint = \sigma\psi_1\psi_2^* + \sigma^*\psi_1^*\psi_2,
\end{gather}
where $\sigma = \const$. In this case, \Eqs{eq:response} yields
\begin{gather}
\oper{D}_1\psi_1 = \sigma^*\psi_2, \quad
\oper{D}_2\psi_2 = \sigma\psi_1;
\end{gather}
hence a quadratic equation for the eigenfrequencies $\freq$,
\begin{gather}\label{eq:lincoupl}
\rho_1\rho_2(\freq-\nu_1)(\freq-\nu_2)=|\sigma|^2, 
\end{gather}
from which $\Psi = \freq_1\action_1 + \freq_2\action_2$ is obtained.

As a particular case, consider interaction of a particle internal mode having frequency $\Omega$ and action $J=\rho|\psi|^2$ with an external oscillating field $E=\bar{E}e^{-i\omega t}$ having frequency $\omega$ and action $I=\rho_E|E|^2$. Given that the field occupies a volume $V \to \infty$, the frequency shifts $\delta\Omega$ and $\delta\omega$ are infinitesimal, \Eq{eq:lincoupl} yielding
\begin{gather}
\delta\Omega\kpt{.5}\rho = \frac{|\sigma|^2}{(\Omega-\omega)\rho_E}, \quad
\delta\omega\kpt{.5}\rho_E = \frac{|\sigma|^2}{(\omega-\Omega)\rho}.
\end{gather}
Since $\rho_E \propto V$, one has $\delta\Omega\kpt{.5}J \ll \delta\omega\kpt{.5}I$, whereas
\begin{gather}
\delta\omega\kpt{.5}I = \frac{|\sigma|^2}{(\omega-\Omega)\rho}\,|\bar{E}|^2
\end{gather}
is nonvanishing. Then one gets
\begin{gather}\label{eq:psi12}
\Psi = \Omega J + \Phi_0, \quad \Phi_0 = - \frac{1}{4}\,\alpha |\bar{E}|^2,
\end{gather}
where $\Phi_0$ is the so-called ponderomotive potential (for the general expression see \App{app:ponder}), an insignificant constant $\omega I$ is removed, and $\alpha = 4|\sigma|^2[(\Omega-\omega)\rho\kpt{.2}]^{-1}$. Since
\begin{gather}\label{eq:singphi}
\Phi_0 = \frac{\kappa_\alpha^2|\bar{E}|^2}{\omega-\Omega},
\end{gather}
where $\kappa_\alpha^2 \equiv  {|\sigma|^2/\rho>0}$, the effective potential becomes infinite at the linear resonance [\Fig{fig:ponder}(a)]. However, nonlinear effects remove this singularity, as we show below.

\subsection{Nonlinear oscillator} 

\begin{figure}
\centering
\includegraphics[width=0.48 \textwidth]{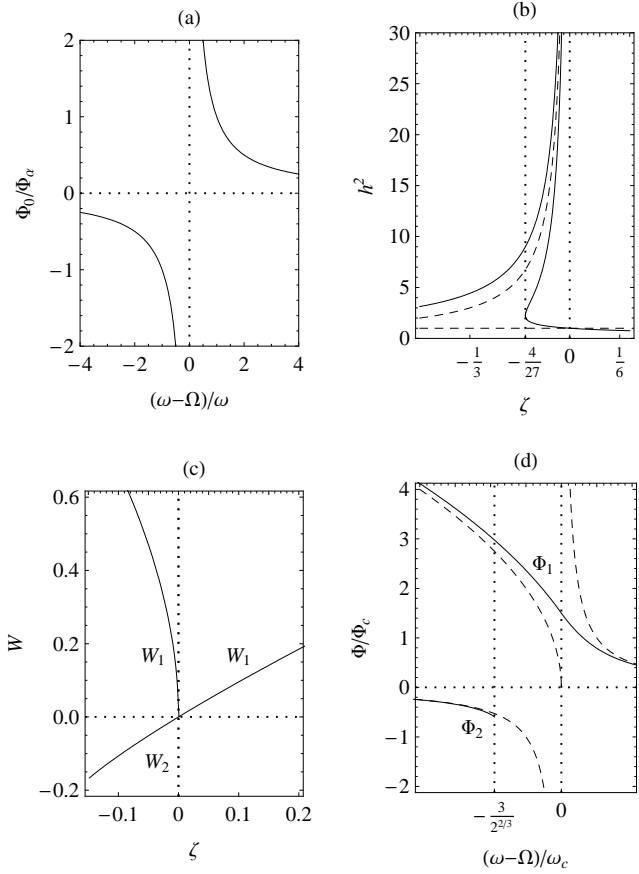}
\caption{(a)~Resonant ponderomotive potential $\Phi_0$ [\Eq{eq:singphi}] in units $\Phi_\alpha\equiv\kappa_\alpha^2|\bar{E}|^2/\omega$ vs. $\omega-\Omega$ in units $\omega$. (b)~Solid -- squared normalized frequency shift $h = \delta\omega I/\Phi_0$ vs. $\zeta = \beta \Phi_0/(\gamma\rho)^2$, $\zeta (J=0) \propto {(\omega-\Omega)^{-3}}$; dashed -- approximations $h^2 = 1$ and $h^2 = -\zeta^{-1}$. (c)~$W(\zeta)$ [\Eq{eq:hystphi}], consists of branches $W_1>0$ and $W_2<0$. (d)~Effective potential $\Phi$ in units $\Phi_c \equiv \rho^2\omega_c^2/\beta = \frac{2}{3}\,\Phi(0)$ vs. $\omega-\Omega$ in units $\omega_c \equiv \beta^{1\!/3}|\sigma\bar{E}|^{2/3}/\rho$, assuming $J=0$ and $\beta > 0$: solid -- \Eq{eq:hystphi} (consists of branches $\Phi_1>0$ and $\Phi_2<0$), dashed -- \Eqs{eq:hystphiapp}.}
\label{fig:ponder}
\end{figure}

Consider the effective potential near a nonlinear resonance, with a Duffing oscillator as a model system. Then
\begin{gather}\label{eq:lintnlin}
\lint = \sigma\psi E^*+\sigma^*\psi^* E + \frac{1}{2}\,\beta|\psi|^4,
\end{gather}
where $\beta=\const$, yielding
\begin{gather}
\oper{D}\psi = \sigma^*E+\beta|\psi|^2\psi, \quad
\oper{D}_EE = \sigma\psi.
\end{gather}
Separate the driven motion from free oscillations, $\psi = X e^{-i(\omega + \delta\omega)t} + Y e^{-i(\Omega + \delta\Omega)t}$, so
\begin{gather}
-\delta\omega\, \rho_E \bar{E} = \sigma X, \\
(\Omega-\omega)\rho X = \sigma^*\bar{E} + 2\beta |Y|^2X + \beta |X|^2X, \\
-\delta\Omega\,\rho Y = \beta |Y|^2Y + 2\beta |X|^2Y.\label{eq:y}
\end{gather}
From \Eq{eq:y}, it follows that $\delta\Omega J \sim \beta|XY|^2$, which we assume, for simplicity, small compared to $\Phi_0 \sim \delta\omega I$; hence $\Psi \approx {\Omega J + \Phi}$, where $\Phi = \int^I_0 \delta\omega\,dI$ is the modified ponderomotive potential. The field frequency shift is found from the cubic equation $(1+\zeta h^2)h=1$ for $h=\delta\omega I/\Phi_0$; $\zeta = \beta \Phi_0/(\gamma \rho)^2$, $\Phi_0 = |\sigma\bar{E}|^2/(\gamma\rho)$, $\gamma = \omega-\Omega+\varsigma(J)$, $\varsigma = 2\beta J/\rho^2$, and the above condition of negligible $\delta\Omega J$ reads $\varsigma/(\omega-\Omega) \ll 1$. Thus,
\begin{gather} \label{eq:hystphi}
\Phi = \frac{\gamma^2\rho^2}{\beta}\,W(\zeta), \quad 
W(\zeta)=\int^\zeta_0 \!h(\tilde{\zeta})\,d\kpt{-.3}\tilde{\zeta},
\end{gather}
where $h(\zeta)$ has three branches [\Fig{fig:ponder}(b)]. One of those is unstable \cite{book:bogoliubov}; hence two branches of the ponderomotive potential, $\Phi_1$ and $\Phi_2$. Their asymptotic form flows from $W_{1,2}(\zeta\to 0\pm) \approx \zeta$ and $W_1(\zeta\to 0-) \approx 2\sqrt{-\zeta}$\kpt{1} [\Fig{fig:ponder}(c)]:
\begin{gather}\label{eq:hystphiapp}
\Phi_{1,2}^\rind{\pm} \approx \Phi_0=\frac{\kappa_\alpha^2 |\bar{E}|^2}{\gamma},\quad
\Phi_1^\rind{-} \approx 2\kappa_\alpha\rho|\bar{E}|\frac{\sqrt{|\gamma\beta|}}{\beta},
\end{gather}
and \Eq{eq:singphi} is recovered from $\Phi_{1,2}^\rind{\pm}$ in the limit of small $\zeta$ and $\varsigma = 0$ [\Fig{fig:ponder}(d)]. On the other hand, $W_1(\zeta\to \infty) \approx \frac{3}{2}\,|\zeta|^{2/3}$, so $\Phi_1 \to 3\kpt{.5}|\sigma\bar{E}|^{4/3}|\beta|^{2/3}\!/(2\beta)$ at the resonance, \ie the singularity vanishes.

\section{Beat resonance}
\label{sec:hyb}

Suppose that $\lint$ also contains a term cubic in $\vec{\psi}$, say,
\begin{gather}
\delta\lint=\epsilon\psi_1\psi_2^*\psi_3^*+\epsilon^*\psi_1^*\psi_2\psi_3,
\end{gather}
where $\epsilon=\const$. That would normally yield a potential small compared to $\Phi_0$, yet maybe except when
\begin{gather}\label{eq:beatr}
\nu_1 \approx \nu_2 + \nu_3.
\end{gather}
Below we study the beat resonance \eq{eq:beatr} neglecting the quadratic $\lint$, so the envelopes $\bar{\psi}_j=\psi_j\kpt{.3} e^{i\nu_{j} t}$ satisfy
\begin{gather}
-i\rho_1\dot{\bar{\psi}}_1 =  \epsilon^*\bar{\psi}_2\bar{\psi}_3e^{-i\Delta t}, \label{eq:hybenv1}\\
-i\rho_2\dot{\bar{\psi}}_2 =  \epsilon \bar{\psi}_1\bar{\psi}_3^*e^{i\Delta t}, \label{eq:hybenv2}\\
-i\rho_3\dot{\bar{\psi}}_3 =  \epsilon \bar{\psi}_1\bar{\psi}_2^*e^{i\Delta t}, \label{eq:hybenv3}
\end{gather}
where $\Delta \equiv \nu_3 + \nu_2 - \nu_1$ is the detuning frequency.

\subsection{Linear coupling}

Suppose that $\Delta$ is large enough, so \Eqsd{eq:hybenv1}{eq:hybenv3} can be solved by averaging. Then split $\bar{\psi}_j$ into driven and free motion: $\bar{\psi}_j = X_j\, e^{-i\eta_j t} + Y_j\, e^{-i\dfreq_{j} t}$, where $\dfreq_{j} \ll \eta_j$ are the nonlinear frequency shifts, and 
\begin{gather}
\eta_1=\dfreq_2+\dfreq_3+\Delta, \\
\eta_2=\dfreq_1-\dfreq_3-\Delta, \\
\eta_3=\dfreq_1-\dfreq_2-\Delta.
\end{gather}
Thus the corresponding equations read
\begin{gather}
\begin{align}
-\eta_1\rho_1 X_1 = \epsilon^* Y_2Y_3, \quad &
-\dfreq_1\rho_1 Y_1 = \epsilon^*\!\!\left(Y_2X_3 +X_2Y_3\right),\kpt{10}\nonumber\\
-\eta_2 \rho_2 X_2 = \epsilon Y_1Y_3^*, \quad &
-\dfreq_2 \rho_2 Y_2 = \epsilon\!\left(Y_1X_3^* +X_1Y_3^*\right),\nonumber\\
-\eta_3 \rho_3 X_3 = \epsilon Y_1Y_2^*, \quad &
-\dfreq_3 \rho_3 Y_3 = \epsilon\!\left(Y_1X_2^* +X_1Y_2^*\right),\nonumber
\end{align}
\end{gather}
where insignificant oscillating terms are neglected. Since $\dfreq_{j}\ll\Delta$, one has
\begin{gather}
\dfreq_1=-\frac{\action_2+\action_3}{\Gamma\Delta}, \quad
\dfreq_2=\frac{\action_3-\action_1}{\Gamma\Delta}, \quad
\dfreq_3=\frac{\action_2-\action_1}{\Gamma\Delta},\nonumber
\end{gather}
where $\Gamma \equiv \rho_1\rho_2\rho_3/|\epsilon|^2 > 0$. Then, from \Eqs{eq:psi}, one gets $\Psi=\Psi_0+\Phi$, 
\begin{gather}\label{eq:dphi0}
\Phi = \frac{\action_1\action_2+\action_1\action_3- \action_2\action_3}{(\nu_1-\nu_2-\nu_3)\,\Gamma}.
\end{gather}

Particularly, if the third mode corresponds to a macroscopic oscillating field tuned close to the beat resonance
\begin{gather}
\omega \approx \Omega_1-\Omega_2,\label{eq:beatres}
\end{gather}
then $J_{1,2} \ll I \propto |\bar{E}|^2$. Hence the ``hybrid'' ponderomotive potential that is obtained is simultaneously proportional to $|\bar{E}|^2$ and the internal actions $J_j$:
\begin{gather}\label{eq:dphi}
\Phi =\frac{\kappa_\epsilon^2}{\Delta}\left(J_2-J_1\right)|\bar{E}|^2,
\end{gather}
where $\Delta = \omega - (\Omega_1-\Omega_2)$, and $\kappa_\epsilon^2 \equiv |\epsilon|^2/(\rho_1\rho_2) > 0$.

\subsection{Nonlinear coupling}

At $\Delta\lesssim\dfreq_{j}$, the internal mode equations
\begin{gather}\label{eq:beatfield}
-i\rho_1\dot{\bar{\psi}}_1 = \epsilon^*\bar{\psi}_2\bar{E}e^{-i\Delta t}, \quad
-i\rho_2\dot{\bar{\psi}}_2 = \epsilon \bar{\psi}_1\bar{E}^*e^{i\Delta t}
\end{gather}
do not allow averaging, so the above derivation is modified as follows. First, consider strictly periodic oscillations, in which case $\bar{\vec{\psi}}\equiv(\bar{\psi}_1,\bar{\psi}_2)\transp$ rewrites as $\vec{\bar{\psi}}= \oper{T}_{\Delta}\kpt{1}\oper{C}^{-1}_\psi\kpt{.3}\vec{x}$, where $\oper{T}_y =\text{diag}(e^{-iyt/2},e^{iyt/2})$, and $\oper{C}_\psi= \oper{T}_{\kpt{-.5}-\phi}\kpt{2}\text{diag} (\rho_1, \rho_2)^{\kpt{-.5}1\kpt{-.2}/2}$, $\phi = \arg(\epsilon\bar{E}^*)+\pi$. Then
\begin{gather}\label{eq:z}
i\dot{x}_1 = \frac{\varepsilon}{2}\,x_2-\frac{\Delta}{2}\,x_1, \quad
i\dot{x}_2 = \frac{\varepsilon}{2}\,x_1+\frac{\Delta}{2}\,x_2,
\end{gather}
where $\varepsilon = 2\kappa_\epsilon|\bar{E}|$. \Eqs{eq:z} yield two eigenmodes at frequencies $\pm\frac{1}{2}\,\Lambda$, $\Lambda = (\varepsilon^2+\Delta^2)^{1/2}$, so $\vec{x} = \oper{U} \oper{T}_{\Lambda}\kpt{.3}\bar{\vec{x}}$, where
\begin{gather}
\oper{U} = \twobytwomatrix{\cos\Theta}{-\sin\Theta}{\sin\Theta}{\cos\Theta},
\end{gather}
$\bar{\vec{x}} \equiv (\bar{x}_\rind{+},\bar{x}_\rind{-})\transp$ is constant, and $\Theta$ satisfies
\begin{gather}
\cos 2\Theta = -\Delta/\Lambda, \quad \sin 2\Theta = \varepsilon/\Lambda. 
\end{gather}
Then
\begin{gather} \label{eq:psiz}
\vec{\psi}= \oper{T}_\omega\oper{S}\kpt{.5} \vec{\chi}, 
\end{gather}
where $\oper{S}\equiv\oper{C}^{-1}_\psi\oper{U}\kpt{.3}\oper{C}_\chi$ is a constant matrix, $\oper{C}_\chi = \text{diag}(\rho_\rind{+},\rho_\rind{-})^{\kpt{-.5}1\kpt{-.2}/2}$, $\rho_\rind{\pm} \equiv \freq_\rind{\pm}$ (for uniformity),
\begin{gather}\label{eq:freqs}
\freq_\rind{\pm} = \frac{1}{2}\,(\Omega_1+\Omega_2) \pm \frac{1}{2}\,\Lambda,
\end{gather}
$\chi_\rind{\pm} = \bar{\chi}_\rind{\pm}\kpt{.3} e^{-i\vartheta_\rind{\pm}}$, $\dot{\vartheta}_\rind{\pm} = \freq_\rind{\pm}$, and $\bar{\vec{\chi}} =\oper{C}_\chi^{-1} \bar{\vec{x}}$. 

When the oscillation parameters evolve, treat \Eq{eq:psiz} as a formal change of variables for \Eq{eq:compagr}. Hence $\lmode = L_\rind{+} + L_\rind{-}$, the field part being omitted, and
\begin{gather}
L_\rind{\pm} = \frac{i\rho_\rind{\pm}}{2}\,
(\dot{\chi}_\rind{\pm}\chi^*_\rind{\pm}-\dot{\chi}^*_\rind{\pm}\chi_\rind{\pm})
-\rho_\rind{\pm}\nu_\rind{\pm}|\chi_\rind{\pm}|^2.
\end{gather}
Therefore $\chi_\rind{\pm}$ are independent linear modes with frequencies $\freq_\rind{\pm}$ and conserved actions $J_\rind{\pm} = \rho_\rind{\pm}|\chi_\rind{\pm}|^2$ (\Sec{sec:eigen}), and \Eq{eq:psi} yields
\begin{gather}\label{eq:truepsi}
\Psi = \frac{1}{2}\,(\Omega_1+\Omega_2)\big(J_\rind{+}+J_\rind{-}\big)+\frac{\Lambda}{2}\,\big(J_\rind{+}-J_\rind{-}\big).
\end{gather}
At $\Delta \gg \varepsilon$, $J_\rind{\pm} = J_{1,2}$ for $\Delta < 0$ and $J_\rind{\pm} = J_{2,1}$ for $\Delta > 0$, so \Eq{eq:dphi} is recovered by Taylor expansion of \Eq{eq:truepsi}.

\subsection{Quantum analogy}
\label{sec:rabi}

\begin{figure}
\centering
\includegraphics[width=0.42 \textwidth]{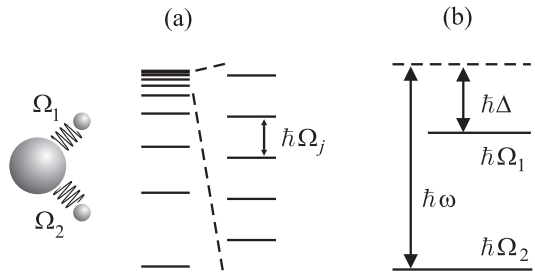}
\caption{(a)~Schematic of a quasiclassical particle containing two harmonic oscillators with unperturbed frequencies $\Omega_1$ and $\Omega_2$. The energy spectrum for each individual oscillator at large energies (close-up on the right) is equidistant. The coupling with an oscillating field $\bar{E}$ near the beat resonance \eq{eq:beatres} is described by \Eqs{eq:beatfield}. (b)~A two-level system, with unperturbed eigenfrequencies $\Omega_1$ and $\Omega_2$, formally described by the same \Eqs{eq:beatfield}, assuming the Rabi frequency $\rabi = \varepsilon$.}
\label{fig:analogy}
\end{figure}

The above classical particle is the limit of a quantum system with plentiful states coupled to the field simultaneously and $\Omega_j$ being the unperturbed transition frequencies [\Fig{fig:analogy}(a)]. Yet, with $\rho_j \to \hbar$, the equations \eq{eq:beatfield} are also equivalent to those describing a \textit{two}-level system  \cite{book:scully97, ref:berman82, ref:courtens77}, with the unperturbed eigenfrequencies $\Omega_1$ and $\Omega_2$ and the Rabi frequency $\rabi = \varepsilon$ [\Fig{fig:analogy}(b)]. Hence \Eq{eq:truepsi} yields as well the dipole potential for a two-level quantum object, \eg a cold atom \cite[pp.~454-461]{book:cohentannoudji}, \cite{ref:dalibard85}:
\begin{gather}\label{eq:rabipsi}
\Psi = \frac{\hbar}{2}\,(\Omega_1+\Omega_2)+ \frac{\hbar}{2}\, \big(n_\rind{+}-n_\rind{-}\big)\sqrt{\Delta^2+\rabi^2},
\end{gather}
where $n_j=|\chi_j|^2$ are the occupation numbers ($J_j \to \hbar n_j$), satisfying ${n_\rind{+}+n_\rind{-}=1}$. Similarly, \Eq{eq:dphi} is equivalent to the dipole potential at weak coupling \cite{ref:ashkin78, book:scully97}:
\begin{gather}\label{eq:rabiphi}
\Phi = \frac{\hbar\rabi^2}{4\Delta}\left(1-2n_1\right).
\end{gather}
[At $n_1=0$, \Eq{eq:rabiphi} is further analogous to \Eq{eq:singphi}, with the resonance at the transition frequency $\Omega_1-\Omega_2$.]

This parallelism originates from \Eq{eq:response} yielding Schr\"odinger-like equations for $\lint$  quadratic in internal mode amplitudes. Therefore, for any classical particle governed by a hybrid potential, an analogue is possible which is governed by a quantum dipole potential, and vice versa. Hence the two types of objects permit uniform manipulation techniques, as also discussed in \Sec{sec:oneway}. 

\section{One-way walls}
\label{sec:oneway}

As an \textit{effective} potential, $\Psi$ can have properties distinguishing it from true potentials. Particularly, it can yield asymmetric barriers, or one-way walls \cite{my:cdlarge, my:ratchet, ref:raizen05, ref:ruschhaupt04, ref:ruschhaupt06, ref:ruschhaupt06c, ref:price07, ref:price08, ref:thorn08}, allowing current drive \cite{ref:suvorov88, ref:litvak93, my:cdprl} and translational cooling \cite{ref:raizen05, ref:dudarev05, ref:ruschhaupt06b, ref:price08}. We explain these barriers as follows.

\begin{figure}
\centering
\includegraphics[width=0.42 \textwidth]{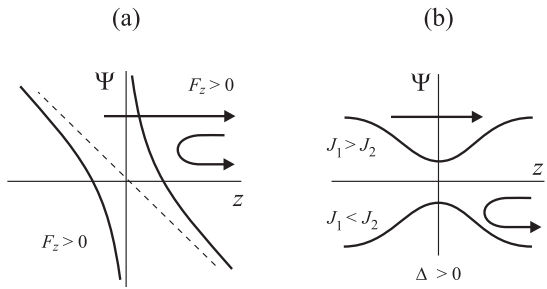}
\caption{Schematic showing two types of one-way walls (arrows denote particle transmission and reflection). (a)~The one-way wall is due to a ponderomotive potential near a primary resonance [\Eqsc{eq:psi12}{eq:singphi}] with uniform $\partial_z\Omega<0$, so $F_z =-\partial_z\Psi > 0$ for all $z$, except at the exact resonance where the effective potential does not apply. [The inclined asymptote corresponds to $\Psi = \Omega J + \const$.] (b)~The one-way wall is due to a hybrid potential [\Eqsc{eq:dphi}{eq:truepsi}] with uniform $\Omega_j$ and $\Delta > 0$: particles incident from the left receive $J_1>J_2$, so they see an attractive $\Psi$; those incident from the right have $J_1<J_2$, so they see a repulsive~$\Psi$.}
\label{fig:wall}
\end{figure}

Suppose a ponderomotive potential of the form \eq{eq:singphi} [or, similarly, \eq{eq:hystphi}]. Given $\Omega=\Omega(z)$, with, say, $\partial_z\Omega<0$, the average force $F_z =-\partial_z\Psi$ is everywhere in $+z$ direction, except at the exact resonance where the effective potential does not apply [\Fig{fig:wall}(a)]. Hence particles can be transmitted when traveling in one direction but reflected otherwise, even assuming uniform $\bar{E}(z)$ \cite{ref:suvorov88, ref:litvak93, my:cdprl}. In \Ref{my:cdlarge}, such dynamics was confirmed for cyclotron-resonant rf fields, and, in \Ref{my:ratchet}, a similar scheme employing abrupt $\bar{E}(z)$ was proposed.

Hybrid potentials (\Sec{sec:hyb}) permit yet another type of one-way walls. Assume uniform $\Omega_j$ and $\Delta > 0$; then $\Psi$ [\Eqsc{eq:dphi}{eq:truepsi}] is repulsive for cold particles ($J_1<J_2$) but attractive for hot particles ($J_1>J_2$). Thus, if particles incident, say, from the left are preheated (via nonadiabatic interaction with another field), they will be transmitted, whereas those cold as incident from the right will be repelled [\Fig{fig:wall}(b)]; hence the asymmetry.

In agreement with the parallelism shown in \Sec{sec:rabi}, similar one-way walls for cold atoms have been suggested \cite{ref:raizen05, ref:ruschhaupt04, ref:ruschhaupt06, ref:ruschhaupt06c, ref:price07, arX:narevicius08} and enjoyed experimental verification \cite{ref:price08, ref:thorn08}. However, the new result here is that quasiclassical particles like Rydberg atoms and molecules can, in principle, be manipulated in the same manner, despite their involved eigenspectrum is different (\Fig{fig:analogy}).

\section{Conclusions} 
\label{sec:conclusions}

We propose a method to calculate the generalized effective potential $\Psi$ for a nonrelativistic classical particle undergoing arbitrary oscillations in high-frequency or static fields. We derive $\Psi$ from the oscillation eigenfrequencies in the particle-field system [\Eq{eq:psi}], like in the dressed-atom approach \cite{book:cohentannoudji, ref:dalibard85, ref:berman82, ref:courtens77} but from nonlinear classical equations [\Eqs{eq:response}]. Specifically, we show that the ponderomotive potential [\Eqsc{eq:singphi}{eq:genphi}; \Fig{fig:ponder}(a)] extended to a nonlinear oscillator has multiple branches near the primary resonance [\Eq{eq:hystphi}; \Fig{fig:ponder}(d)]. Also, for a pair of natural frequencies in a beat resonance, $\Psi$ scales linearly with the internal actions [\Eqsc{eq:dphi}{eq:truepsi}] and is analogous to the dipole potential [\Eqsc{eq:rabipsi}{eq:rabiphi}] for a two-level quantum system (\Fig{fig:analogy}).  Thus cold quantum particles and highly-excited quasiclassical objects permit uniform manipulation tools, particularly, stationary asymmetric barriers, or one-way walls~(\Fig{fig:wall}). 

\section{Acknowledgments}

This work was supported by DOE Contracts No.~DE-FG02-06ER54851, DE-FG02-05ER54838, and by the NNSA under the SSAA Program through DOE Research Grant No.~DE-FG52-04NA00139.

\appendix

\section{General expression for the ponderomotive potential}
\label{app:ponder}

Consider generalization of the ponderomotive potential \eq{eq:psi12}, \eq{eq:singphi} to multiple internal oscillations and vector field
\begin{gather}
\bar{\vec{E}} = \sum_\mu \bar{\vec{E}}_\mu, \quad 
\bar{\vec{E}}_\mu= \vec{e}_\mu \bar{E}_\mu,
\end{gather}
composed of modes $\mu$ with polarizations $\vec{e}_\mu$. From \Sec{sec:pondlin}, it is known that the internal energy $\delta\vec{\Omega} \cdot \vec{J}$ yields a negligible contribution to $\Psi$. Thus
\begin{gather}
\Phi_0 = \sum_\mu \delta\omega_\mu I_\mu,
\end{gather}
where the infinitesimal frequency shifts $\delta\omega_\mu$ of the field are found as follows. Use \cite{my:dipole}
\begin{gather}
\lint = \frac{1}{2}\,\Re\big[\bar{\vec{E}}^*\cdot\bar{\vec{d}}\big],
\end{gather}
where insignificant oscillating terms are removed, and $\bar{\vec{d}}$ is the particle induced dipole moment. From \Eq{eq:response}, 
\begin{gather}\label{eq:a1}
-\delta\omega_\mu\rho_\mu\bar{E}_\mu = \frac{1}{4}\,\vec{e}_\mu^*\cdot \bar{\vec{d}}.
\end{gather}
Multiply \Eq{eq:a1} by $\bar{E}_\mu^*$ and substitute $\bar{\vec{d}}=\oper{\alpha}\bar{\vec{E}}$, where $\oper{\alpha}$ is the polarizability tensor; then
\begin{gather}
\delta\omega_\mu I_\mu = - \frac{1}{4}\,\bar{\vec{E}}_\mu^*\cdot\oper{\alpha}\bar{\vec{E}}.
\end{gather}
Hence summation over all field modes yields the known expression \cite{ref:auerbach66, ref:gordon73}
\begin{gather}\label{eq:genphi}
\Phi_0 = -\frac{1}{4}\,(\bar{\vec{E}}^*\cdot\oper{\alpha}\bar{\vec{E}}),
\end{gather}
which holds for any internal modes contributing to $\oper{\alpha}$~\cite{my:dipole}.


\begin{thebibliography}{10}
\bibitem{foot:time}
Slow dependence on time is also allowed here and further.
\bibitem{foot:fieldmode}
In addition to electromagnetic modes \cite{ref:flores05}, any periodic forces
  can be included here, assuming that an extended phase space is used
  \cite{my:beatf}.
\bibitem{foot:dipole}
\Sec{sec:eigen} corrects Appendix~B of our \Ref{my:dipole}.
\end{thebibliography}
\end{document}